\documentclass[twocolumn,twoside,slac_two]{revtex4}
\usepackage{graphicx}
\usepackage{fancyhdr}
\usepackage{amsmath, amssymb}
\begin{document}

\title{Charmonium from Lattice QCD}

%

\author{J.J. Dudek}
\affiliation{Jefferson Lab, 12000 Jefferson Avenue, Newport News, VA 23606, USA \\
Department of Physics, Old Dominion University, Norfolk, VA 23529, USA}

\begin{abstract}
Charmonium is an attractive system for the application of lattice QCD
methods. While the sub-threshold spectrum has been
considered in some detail in previous works, it is only very recently
that excited and higher-spin states and further properties such as
radiative transitions and two-photon decays have come to be
calculated. I report on this recent progress with reference to work
done at Jefferson Lab.
\end{abstract}

\maketitle

\thispagestyle{fancy}


\section{Introduction}
Between $3$ and $3.7$ GeV a number of states exist which are believed
to be the bound states of a charm quark and an anti-charm quark and
whose widths are narrow owing to their being below the threshold to decay
to a pair of open charm mesons coupled with the suppression of
annihilation channels at this high mass scale. Because the hadronic
contributions to their widths are so small, radiative transitions
between them constitute considerable branching fractions, and the
rates of these transitions have been measured with some accuracy by a
number of experiments (\cite{Yao:2006px}). Additionally the $C=+$ states can
decay to a pair of photons - this process when time-reversed can serve
as a production mechanism (two-photon fusion) at $e^+ e^-$ machines.

Rates for these radiative processes have been computed in various
varieties of quark-model, and are typically fairly successful when one
sets parameters using the experimental spectrum, however corrections beyond
approximations like non-relativistic dynamics are often uncontrolled
in these models.

In the current century the charmonium picture has filled out
considerably and new mysteries have arisen owing to the high
statistics and new production methods made possible by CLEO-c and the
$B$-factories. The remaining expected sub-threshold states, $\eta_c',\,
h_c$, have been observed, as have radiative transitions from the
$\psi(3770)$ down to the $\chi_{cJ}$. The above-threshold spectrum is
rapidly being mapped (\cite{Swanson:2006st}), with some states living up to the expectations
of potential models (\cite{Uehara:2005qd}) and others coming as something of a
surprise (\cite{Choi:2003ue}). The increasingly complete set of exclusive data
in $e^+ e^-$ looks set to allow determination of the vector
spectrum with some confidence. 

In a series of recent works (\cite{Dudek:2006ej, Dudek:2006ut,
  Dudek:2007wv}), members of the Jefferson Lab lattice group have
investigated the possibility of computing excited spectral and radiative quantities
using lattice QCD. These initial studies have been carried out on
quenched lattices with rather promising results. In the sections that
follow I will briefly describe the work done.

\section{Excited and higher spin states}
The mass spectrum of a field theory considered in Euclidean space-time
can be extracted from the time-dependence of a two-point correlation
function,
\begin{equation}
  C_{ij}(t) = \sum_{\vec{x}} \langle {\cal O}_i(\vec{x},t) {\cal
    O}_j(\vec{0},0)\rangle,\nonumber
\end{equation}
where ${\cal O}_{i,j}$ are operators that have the right quantum numbers to produce a
particular state from the vacuum which are constructed from the fundamental
fields of theory. For example in QCD we might try to
study the pseudoscalar spectrum by considering an operator $\bar{\psi}
\gamma^5 \psi$. The correlator receives contributions from all states
in the theory with the appropriate quantum numbers,
\begin{equation}
   C_{ij}(t) = \sum_\alpha \frac{Z^{\alpha*}_i Z^{\alpha}_j}{2 m_\alpha} \exp  -
     m_\alpha t .\label{corr}
\end{equation}
(In a quantum mechanical bound state model we might think of radial
excitations being labeled by $\alpha$). In practice extracting anything other than the ground state mass from
fits to the time-dependence of a single correlator is difficult and often unstable. This is
particularly troublesome in a system like charmonium where there are
significant approximate degeneracies, e.g. the $\psi(3686)$ and
$\psi(3770)$. These degeneracy problems are made worse on a cubic
lattice where states are labeled not by a continuum spin, but by an
irreducible representation of the cubic group. The continuum spin
content of these various irreps is shown in Table I. This indicates
that, for example, componets of a $3^{--}$ state would appear in the same
correlators as $1^{--}$ states. Since from potential models we expect
there to be a $3^{--}(^3 D_3)$ roughly degenerate with the
$1^{--}(^3D_1)\,\psi(3770)$, we anticipate that there should be
three roughly degenerate excited states above the ground state in a
$T_1^{--}$ correlator. Extracting this from a fit to a single
correlator is not practical.

\begin{table}
\begin{tabular}{ccc}
$\Lambda$ & $d_{\Lambda}$ & $J$\\ \hline
$A_1$ & 1 & $0,4,6,\dots$\\
$A_2$ & 1 & $3,6,7,\dots$\\
$E$ & 2 & $2,4,5,\dots$\\
$T_1$ & 3 & $1,3,4,\dots$\\
$T_2$ & 3 & $2,3,4,\dots$\\ \hline
\end{tabular}
\caption{The table shows the single-valued irreducible representations
  $\Lambda$ of the cubic group $O$, together with their dimensions
  $d_\Lambda$ and continuum spin content $J$\protect.  Additional
  superscripts are employed to denote charge conjugation $C$ and
  parity $P$.\label{tab:irrep}}
\end{table}

Given this one might consider more reliable ways to extract the excited state spectrum. A
variational method utilizing a large basis of operators satisfies this
need. Its major advantage is that it utilizes the orthogonality of
states in a space of operators - while states might be degenerate and
hence hard to separate on the basis of mass, they remain orthogonal
and hence easier to separate on the basis of their state vectors.

In \cite{Dudek:2007wv}, an operator basis was constructed based upon operators that in the
continuum would have the structure of fermion bilinears with a number
of symmetric covariant derivatives
\begin{equation}
{\cal O}_{\mu\nu\rho\cdots} = \bar{\psi}(x) \Gamma_\mu \overleftrightarrow{D}_\nu
\overleftrightarrow{D}_\rho \cdots \psi(x).\nonumber
\end{equation}
Including up to two derivatives, these operators give access to almost
all continuum $J^{PC}$ with $J\leq 3$. Suitable linear combinations of
these operators can be constructed that transform as the irreducible
representations of Table I. These are related to the operators used in
\cite{Liao:2002rj}.

Once a matrix of correlators, $C_{ij}(t)$, has been computed (for a given irrep),
the mass spectrum follows from solution of a generalized eigenvalue
problem that can be shown to be the quantum mechanical variational
solution. We solve
\begin{equation}
C(t) v_\alpha = \lambda_\alpha(t) C(t_0) v_\alpha,\label{geneig}
\end{equation}
for the eigenvalues $\lambda_\alpha(t)$ which are related to state
masses, and for the eigenvectors $v_\alpha$ which are related to the
overlap of our operators onto the mass eigenstates, the $Z^{\alpha}_i$
in eqn (\ref{corr}). 

\begin{figure}
\centering
\includegraphics[width=85mm]{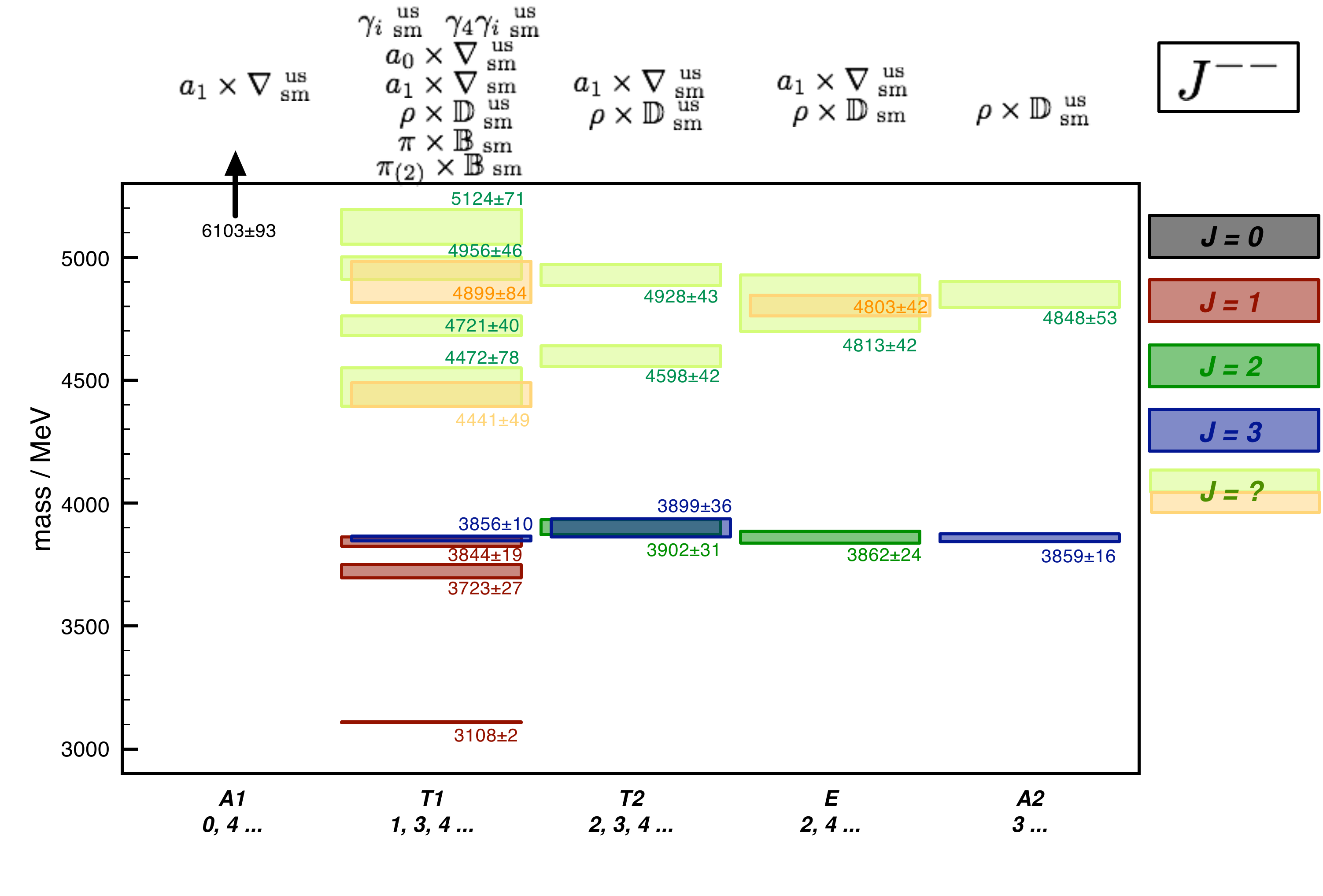}
\caption{Extracted mass spectrum for $PC=--$ listed by lattice
  irreducible representation. Operator labels listed in \cite{Dudek:2007wv}} \label{plot--}
\end{figure}

We computed correlators on quenched anisotropic lattices with $a_s \sim
0.1\,\mathrm{fm}$ and $a_t^{-1} \sim 6\,\mathrm{GeV}$. Full details can
be found in \cite{Dudek:2007wv}.

We show in figure \ref{plot--} the mass spectrum extracted for negative parity
and charge conjugation. In the $T_1$ representation we see precisely
the level structure we expected, namely a ground state and three
closely spaced excited states above. Looking at the states in the
other irreps we see that one possible continuum spin assignment of the
states in the first excited ``band'' would be to have two spin-1 states,
one spin-2 state and one spin-3 state\footnote{In the continuum the
  appearance of a e.g. spin-2 state in both the three-dimensional $T_2$ and
  the two-dimensional $E$ corresponds to the five spin projections of
  a spin-2 meson}. We gain a good deal of support
for this hypothesis from studying the eigenvectors extracted from eqn (\ref{geneig}). Consider the lattice irrep projections of the ``$a_1
\times \nabla$'' operator:
\begin{eqnarray}
  {\cal O}_{T_2}^i = |\epsilon^{ijk}| \bar{\psi}(x) \gamma_5 \gamma_j
  \overleftrightarrow{D}_k\psi(x) \nonumber \\
  {\cal O}_{E}^i = \mathbb{Q}^{ijk} \bar{\psi}(x) \gamma_5 \gamma_j
  \overleftrightarrow{D}_k\psi(x), \nonumber
\end{eqnarray}
where $|\epsilon^{ijk}|, \, \mathbb{Q}^{ijk}$ are Clebsch-Gordan
coefficients for the lattice cubic group. In the continuum we know the
form that the overlap of these operators onto a spin-2 state takes, so
that
\begin{eqnarray}
\langle 0 | {\cal O}_{T_2}^i | 2^{--}(\vec{p}, r) \rangle = Z
|\epsilon^{ijk}| \in_{jk}(\vec{p}, r) \nonumber\\
\langle 0 | {\cal O}_{E}^i | 2^{--}(\vec{p}, r) \rangle = Z
\mathbb{Q}^{ijk}  \in_{jk}(\vec{p}, r),\nonumber
\end{eqnarray}
where $Z$ is common to both. $Z$ can be extracted from the
eigenvectors and if it is found to be close in value in the $T_2$ and
$E$ cases then we conclude that it is likely that we have a spin-2
state.

\begin{figure}
\centering
\includegraphics[width=85mm]{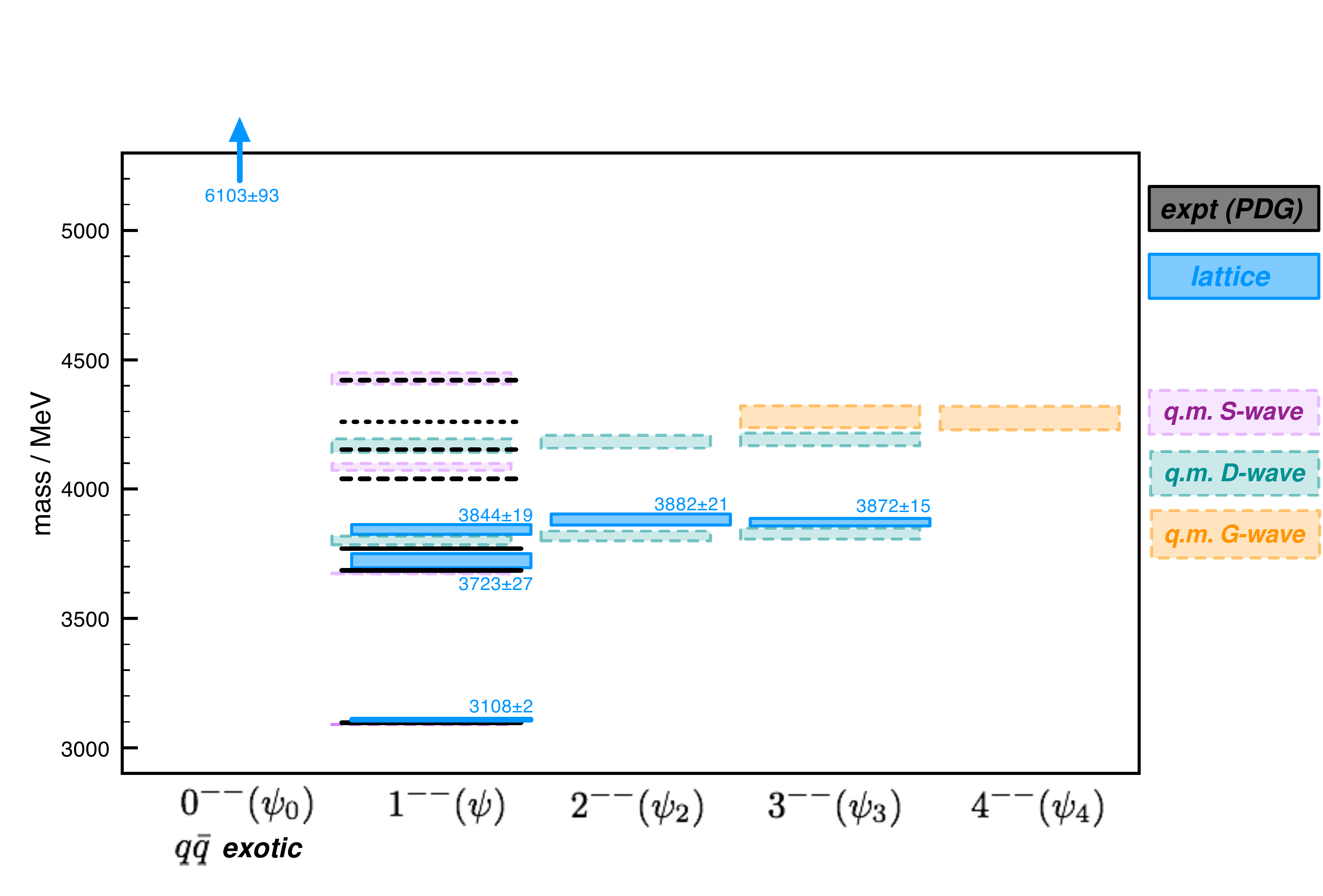}
\caption{Extracted mass spectrum for $PC=--$ listed by assigned
  continuum spin. \label{spectrum--}}
\end{figure}

We apply this eigenvector inspection method wherever possible and
where the result is conclusive we assign the continuum spin shown by the
color coding in figure \ref{plot--}. For the states above the first excited
band this method gave inconclusive answers and for this reason we do
not try to assign a continuum spin. 

This first calculation was performed only at one (quenched) lattice spacing and
consequently our results are not extrapolated to the
continuum. Nevertheless we present our results for continuum spin
assigned states in figure \ref{spectrum--} along with experimental
state masses taken from the PDG(\cite{Yao:2006px}) and potential model masses
taken from \cite{Barnes:2005pb}.

It is clear that we are in agreement with the gross structure
predicted by potential models, and in particular we appear to have
successfully extracted something like the $\psi(3686)/\psi(3770)$
system. We believe that this has not been achieved before in a lattice
calculation. Extracted state masses appear to be systematically high
with respect to potential models and experiment - our suspicion is
that this is due to some combination of computation at finite lattice
spacing and the quenched approximation\footnote{In particular the
  problem of scale setting when one has an incorrect running of the
  coupling. In \cite{Dudek:2007wv} the effect of finite box size was
  tested and was found not to be the source of the level raising effect.} - this hypothesis can be tested with further calculation
now that this method has been demonstrated.

\begin{figure}
\centering
\includegraphics[width=85mm]{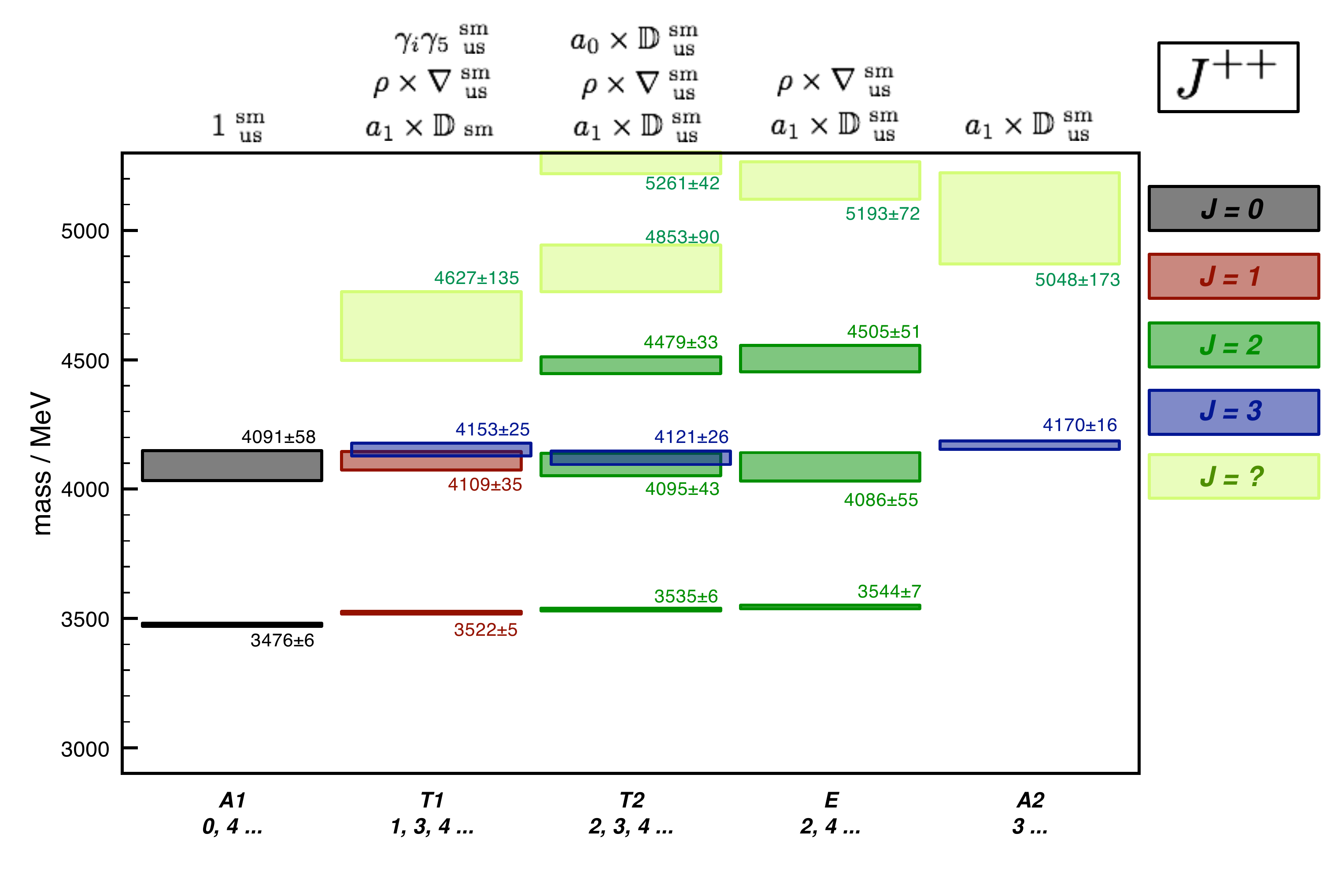}
\includegraphics[width=85mm]{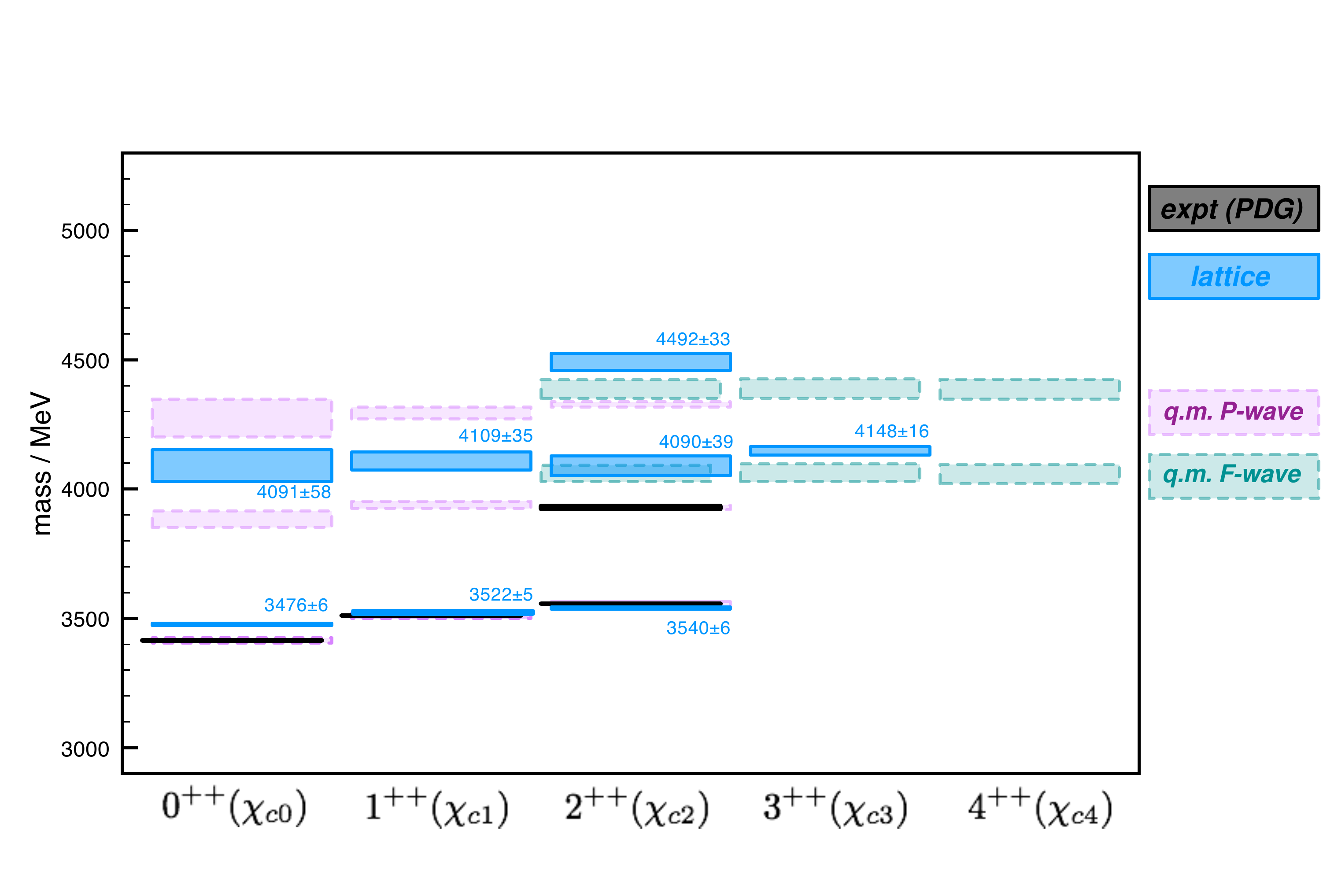}
\caption{Extracted mass spectrum for $PC=++$ listed by lattice
  irreducible representation and continuum spin assigned states. \label{plusplus}}
\end{figure}

Other $PC$ combinations were also considered. In figure \ref{plusplus}
we show our results for $J^{++}$. It is clear that again we are
observing masses systematically higher than the potential model
states. That we miss the spin-4 state near 4 GeV may be
related to the fact that our operators, which have a maximum of two
spatial derivatives do not have any overlap with spin-4 mesons in the
continuum limit. This could be remedied by enlarging the operator basis.

\begin{figure}
\centering
\includegraphics[width=85mm]{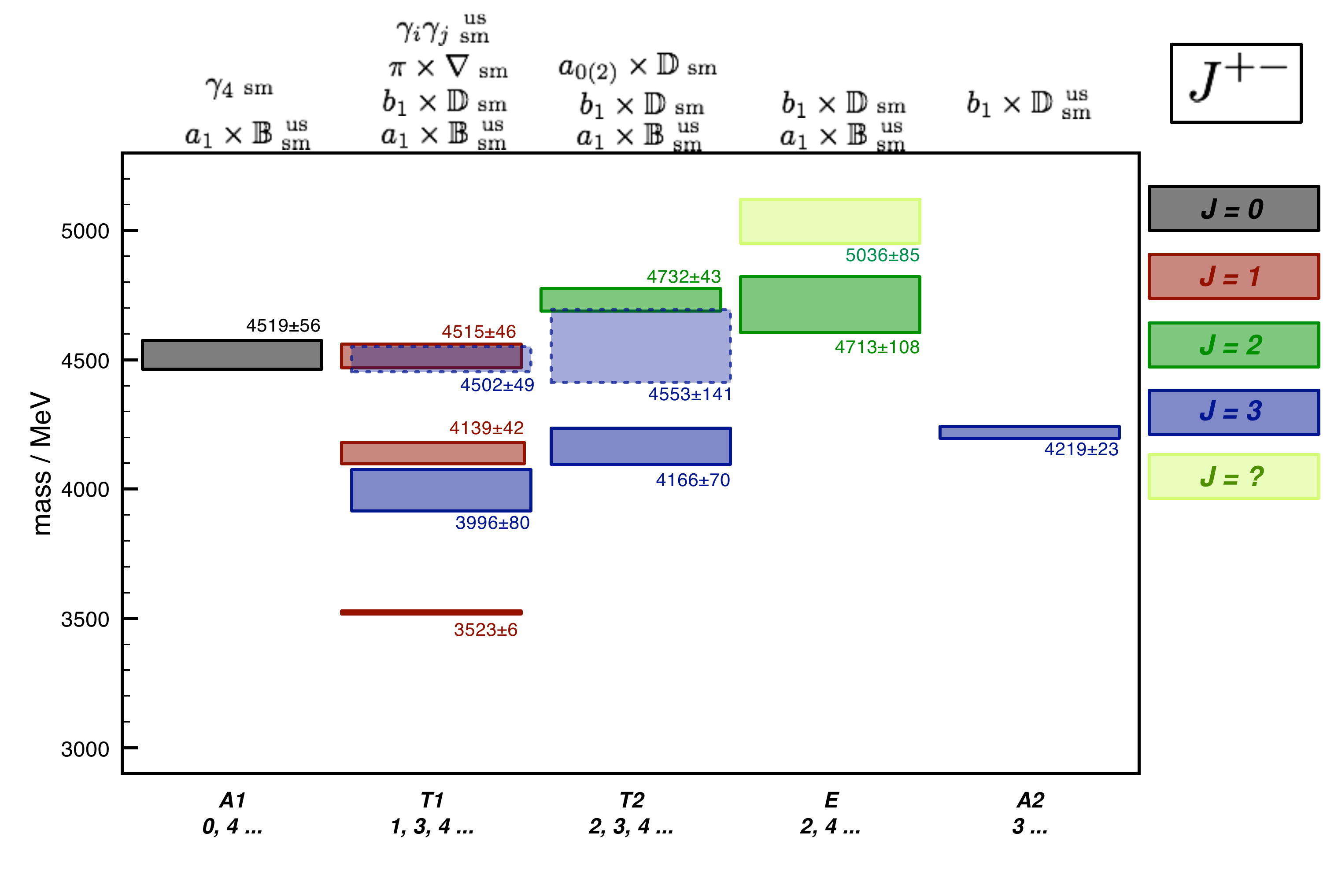}
\includegraphics[width=85mm]{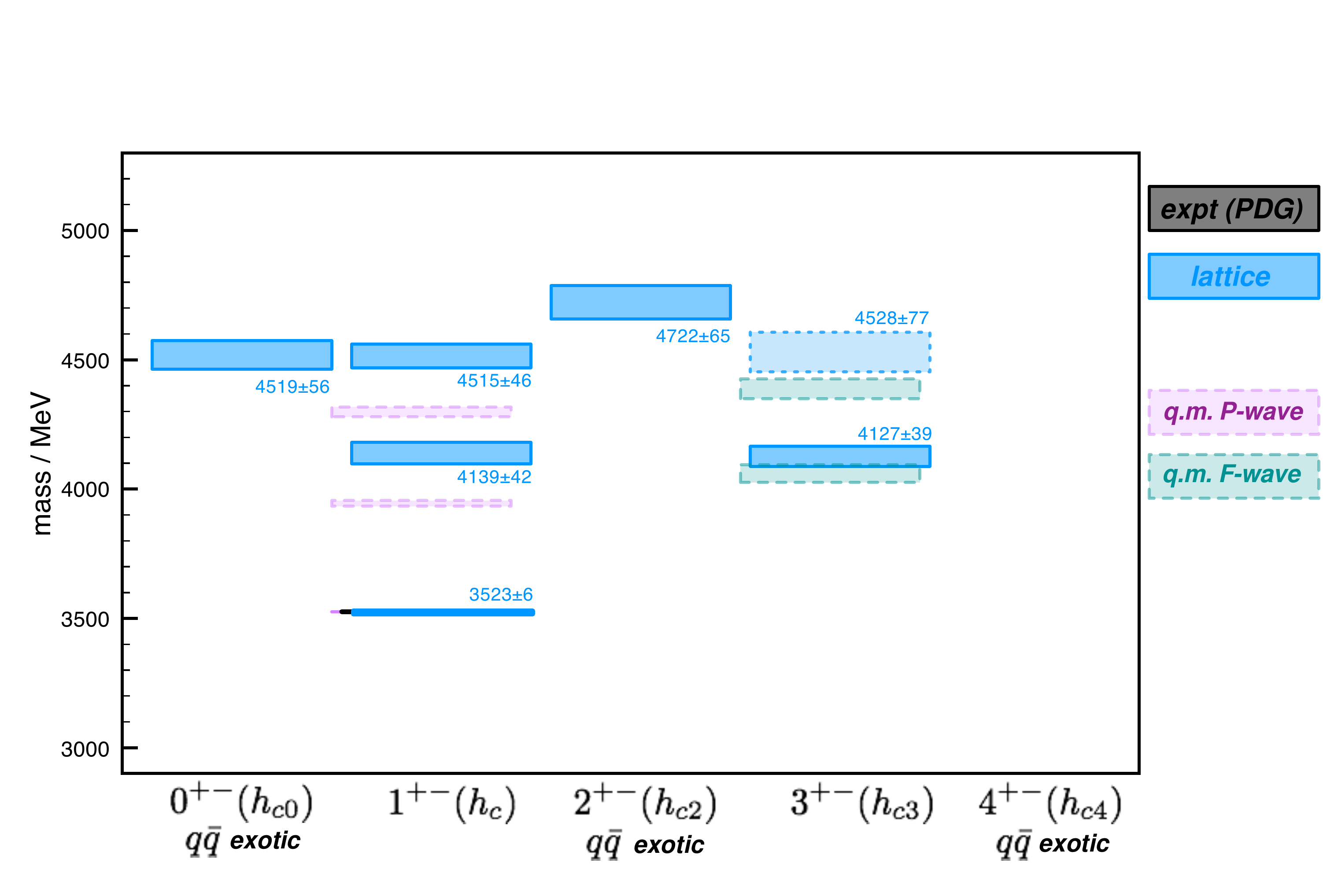}
\caption{Extracted mass spectrum for $PC=+-$ listed by lattice
  irreducible representation and continuum spin assigned states. \label{plusminus}}
\end{figure}

With $PC=+-$, as well as spin-singlets with odd $J$, one also has the
possibility of exotic quantum numbers, i.e. those not accessible to a
$q\bar{q}$ Fock state. In a quenched heavy-quark calculation these can
only arise through non-trivial gluonic excitation giving rise to
states usually described as ``hybrids''. Our extracted mass spectrum
is shown in figure \ref{plusminus} where exotic states with $0^{+-}$,
$2^{+-}$ quantum numbers appear above 4.5 GeV.

\begin{figure}[t]
\centering
\includegraphics[width=85mm]{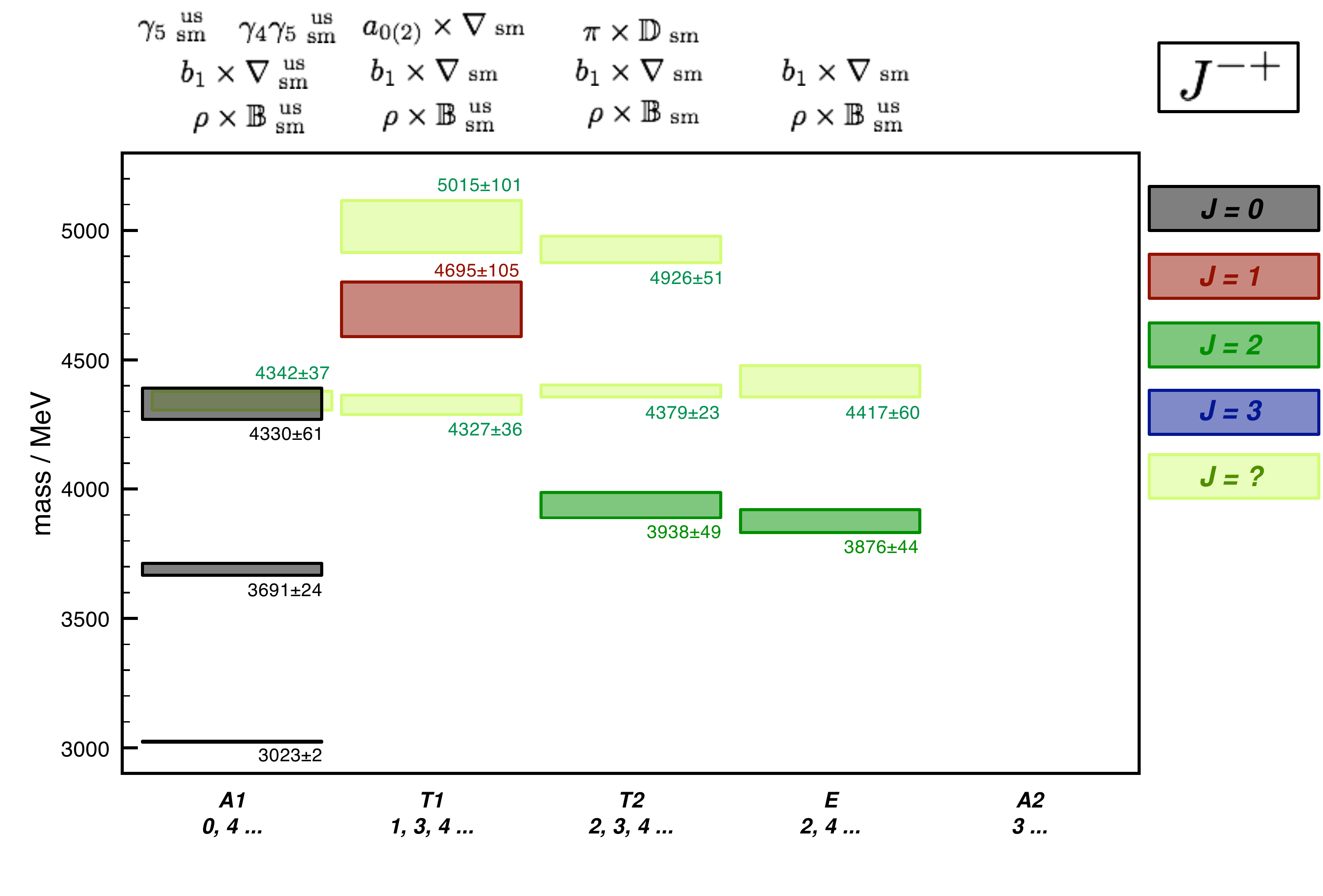}
\caption{Extracted mass spectrum for $PC=-+$ listed by lattice
  irreducible representation. \label{plot-+}}
\end{figure}

\begin{figure}[b]
\centering
\vspace{10cm}
\includegraphics[width=85mm]{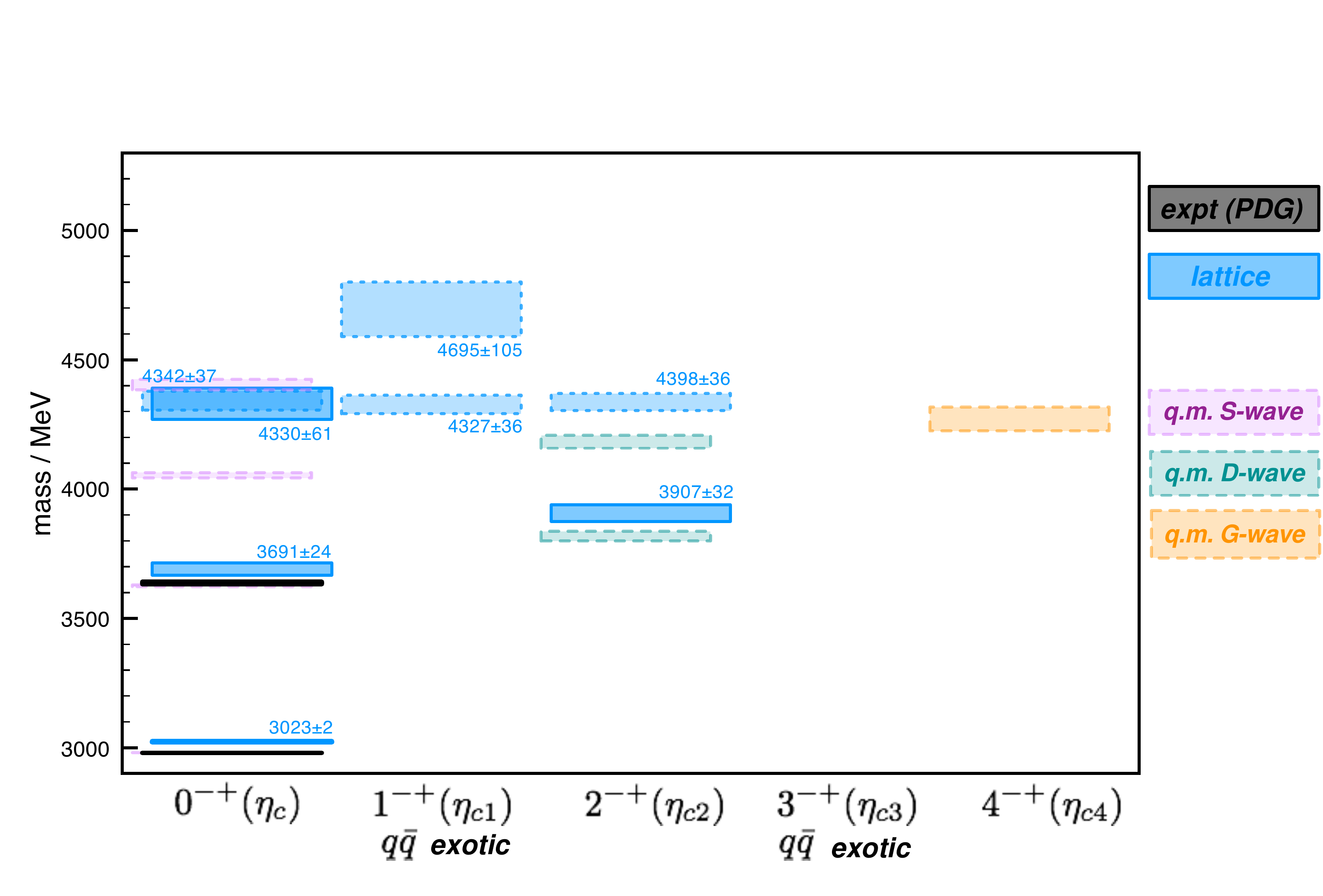}
\includegraphics[width=85mm]{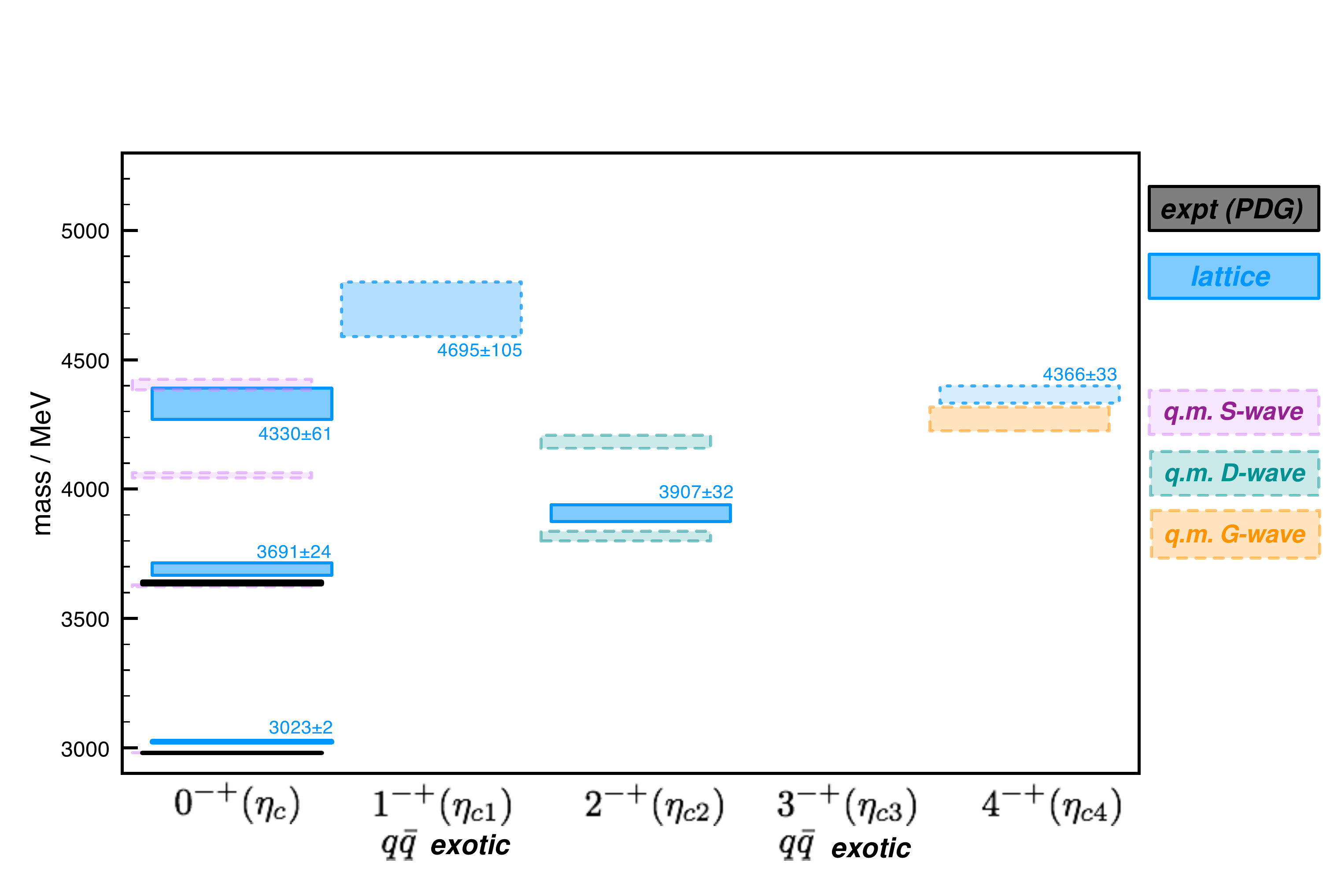}
\caption{Two possible continuum spin interpretations of extracted mass spectrum for $PC=-+$. \label{minusplus}}
\end{figure}

With $PC=-+$, the odd-$J$ states are exotic. Our extracted mass
spectrum listed by lattice irrep is shown in figure \ref{plot-+}. This
case demonstrates the difficulty in continuum spin assignment; the set
of five levels near 4.3 GeV could, on the basis of their mass
degeneracy, be interpreted either as a single $0^{-+}$ and a single
(non-exotic) $4^{-+}$ {\bf or} as two $0^{-+}$ states, an exotic
$1^{-+}$ and a $2^{-+}$. In previous cases we used the eigenvector
inspection method to break these ambiguities, but unfortunately here
the method produces inconclusive results. We display the two possible spectra in figure
\ref{minusplus} where we note that the potential model does have a
$4^{-+}$ state in this mass range. 

It is worth pointing out that previous studies of the
$1^{-+}$ state in charmonia have not taken into account the spin
ambiguity and hence they may have in fact reported the mass of a
non-exotic $4^{-+}$ state. It is clear that further study with more
operators and higher statistics is needed in order to make a definitive
statement.

We believe that we have demonstrated the power of using a variational
solution in a large, carefully constructed operator basis to extract
excited states in lattice QCD. Of course there remain numerous issues
to deal with, including the effect of multiparticle ($D\bar{D}$) states
when one relaxes the quenched approximation, but given that they too
are orthogonal states we should be well-equipped with the method
outlined and an extended basis featuring operators with good overlap
on to these multiparticle states.
 
\clearpage

\section{Radiative transitions}

Sub-$D\bar{D}$-threshold charmonia have very narrow widths such that
radiative transitions between them constitute considerable branching
fractions, these have been measured by a range of experiments and
their relative magnitudes give us clues to internal structure.

These transition widths can be computed using lattice QCD by
considering three-point correlators of the type
\begin{equation}
  C_{i \mu j}(t) \sim  \langle {\cal O}_i(\vec{x},t_f)
  {\cal V}_\mu(\vec{y}, t) {\cal
    O}_j(\vec{0},0)\rangle,\nonumber
\end{equation}
where ${\cal O}_{i,j}$ are operators having overlap with meson states
and ${\cal V}_\mu$ is a lattice representation of the vector
current. For example we could extract the $J/\psi \to \eta_c \gamma$
matrix element from the large Euclidean times value of the correlator
\begin{multline}
  C_{\mu \nu}(t) = \sum_{\vec{x}} e^{-i \vec{p}_f\cdot\vec{x}} e^{i
    \vec{q}\cdot\vec{y}}\\
\times  \Big\langle \big[\bar{\psi} \gamma_5 \psi\big](\vec{x},t_f)
  \big[\bar{\psi}\gamma_\mu\psi\big](\vec{y}, t) \big[\bar{\psi}
  \gamma_\nu \psi\big](\vec{0},0)\Big\rangle.\nonumber
\end{multline}
Correlators of this type, using only point-like operators were evaluated (details can be found in \cite{Dudek:2006ej})
and transition form-factors extracted for a set of transitions between
$J^{PC}$ ground states.

\begin{figure}
\centering
\includegraphics[width=95mm]{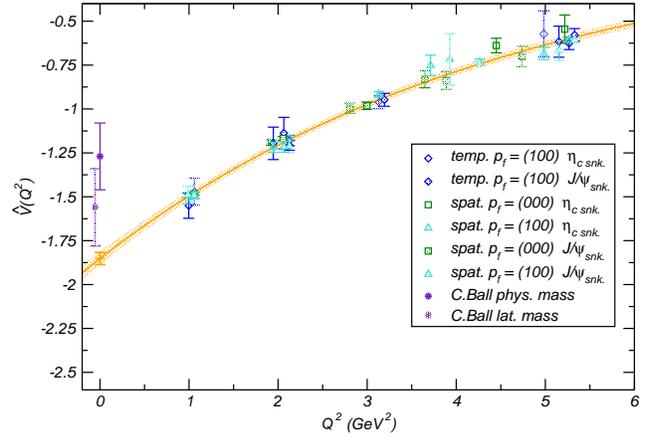}
\caption{Transition form factor for $J/\psi \to \eta_c \gamma$. \label{psieta}}
\end{figure}

The $J/\psi \to \eta_c \gamma$ transition form-factor shown in figure \ref{psieta} is the most
statistically precise signal, but it suffers from a large systematic
issue related to quenching. It is well known that the experimental hyperfine
splitting in charmonium is not reproduced well by studies utilizing
the quenched approximation. As such we have an ambiguity when
computing the phase space that is required to scale a matrix-element
to a width (or vice-versa) - should we use the experimental value or
the value extracted from the spectrum portion of our lattice
calculation? In figure \ref{psieta} we show the experimental
width\footnote{We note that there is ongoing work at CLEO to confirm
  the single measurement from Crystal Ball}
scaled to a matrix-element by both possibilities and the lattice data
fitted with an exponential in photon virtuality, $Q^2$, used to
extrapolate back to $Q^2=0$.

\begin{figure}
\includegraphics[width=95mm]{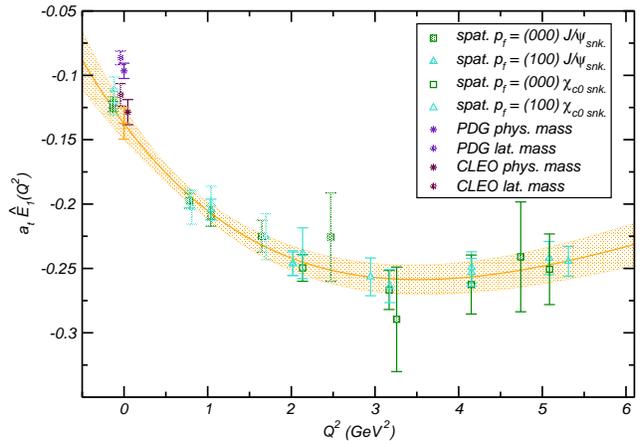}
\caption{Electric dipole transition form factor for $\chi_{c0} \to
  J/\psi \gamma$. Experimental data from PDG(2005) and CLEO(\cite{Adam:2005uh}) \label{chipsi}}
\end{figure}

A transition with reasonable statistical precision and a very small
phase-space ambiguity is the electric dipole transition $\chi_{c0} \to
J/\psi \gamma$. Our results are shown in figure \ref{chipsi} where the
fit uses a form motivated by the quark model. Note the points at
slightly timelike $Q^2$ are not included in the fit - the agreement with
the extrapolated curve then lends support to the fitting form used.

Results for other transitions can be found in \cite{Dudek:2006ej} as can
comparison of the lattice results to potential model
expectations. Work is currently underway combining the excited state technology of the first section
with the radiative transition technology to make it possible to study
transitions involving excited and high-spin states. This would include
experimentally measured transitions like $\psi(3686) \to \chi_{cJ}
\gamma$.

\pagebreak[4]
\section{Two-photon decays}

At first sight it is not clear how one would go about evaluating the
matrix element for the process $\eta_c \to \gamma \gamma$ in lattice
QCD. In the previous section we outlined how to extract the matrix
element for a radiative transition between two QCD eigenstates from
a three-point function evaluated at large Euclidean times. This issue
here is that the photon is not an eigenstate of QCD - taking a vector
interpolating field to large Euclidean time would not yield a photon
state, but instead the lightest QCD vector eigenstate (the $J/\psi$ in this
case).

However, all is not lost, for while the photon is not a QCD
eigenstate, it can be constructed from a linear superposition of QCD
eigenstates. The precise field-theoretic mechanism for this is the LSZ
reduction. The connection in Euclidean space-time, for a different
physical process, is made in \cite{Ji:2001wh} and for the process in
question an outline appears in \cite{Dudek:2006ut}. The end result is that
the following relationship connects the matrix element of interest to a
Euclidean three-point function computable on the lattice:  $\langle
\eta_c(p) | \gamma(q_1, \lambda_1) \gamma(q_2, \lambda_2)\rangle \sim $
\begin{multline}
e^2  \epsilon_\mu(q_1,
 \lambda_1) \epsilon_\nu(q_2, \lambda_2) \int dt_i
e^{- \omega_1 (t_i -t)} \\
\times\Big\langle  \int d^3 \vec{x}\, e^{-i\vec{p}.\vec{x}}
{\cal O} (\vec{x}, t_f) \int d^3 \vec{y}\, e^{i\vec{q_2}.\vec{y}}
{\cal V}^\nu(\vec{y}, t) {\cal V}^\mu(\vec{0}, t_i)       \Big\rangle  \label{master}
\end{multline}

The difference with respect to the radiative transitions between
hadrons considered above is that an integral over the Euclidean time position
of a vector source is now involved.

\begin{figure}
\centering
\includegraphics[width=85mm]{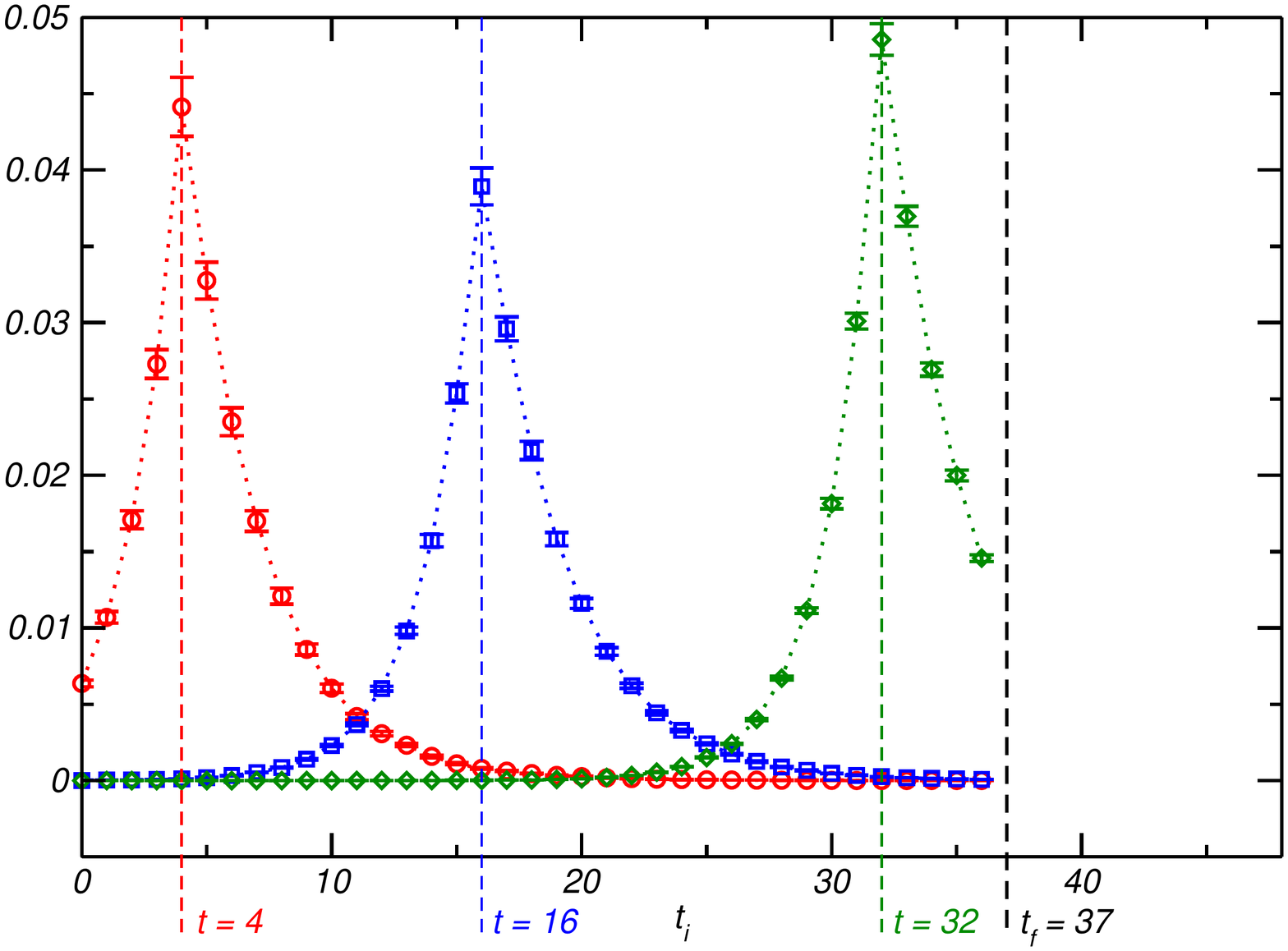}
\includegraphics[width=85mm]{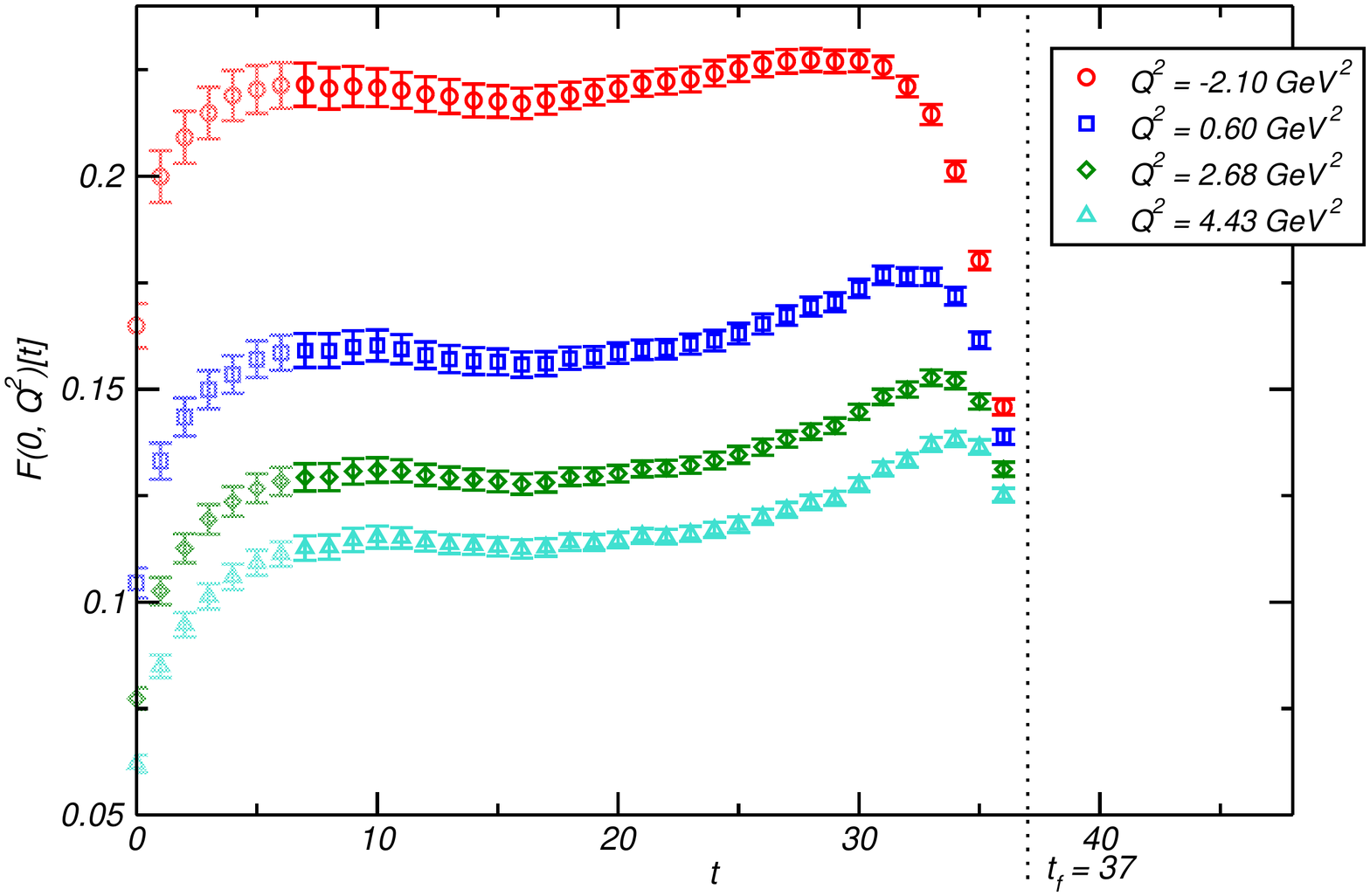}
\caption{\label{int-plat}(a) Integrand in equation\protect{ \ref{master} } at three values
   of vector current insertion time ($t=4,16,32$) with pseudoscalar
   sequential source at sink position $t_f=37$. (b) Pseudoscalar
   two-photon form-factor as a function of time slice, $t$, from equation
   \protect{ \ref{master}}. First six time slices ghosted out due to the Dirichlet wall truncating the integral.} 
\end{figure}

The details of the lattice computation of this object can be found in
\cite{Dudek:2006ut}, here we mention only that an isotropic lattice was used. In figure \ref{int-plat}(a) we display the integrand of equation \ref{master}, having
computed with an operator $\bar{\psi} \gamma_5 \psi$ fixed at $t_f=37$, a conserved
vector current insertion at $t=4,16,32$ and a vector
interpolating field at all possible source positions, $t_i=0 \to 37$. It
is clear that provided one is not too close to the dirichlet wall or
to the sink position, one can capture the entire integral by summing
timeslices. In figure \ref{int-plat}(b) the results of summing timeslices to compute
the integral for all possible insertion positions and a number of
$Q^2$ are shown - clear plateaus are visible at intermediate times
indicating dominance of the $\eta_c$ over the possible excited states.

\begin{figure} 
\begin{center} 
\includegraphics[width=75mm]{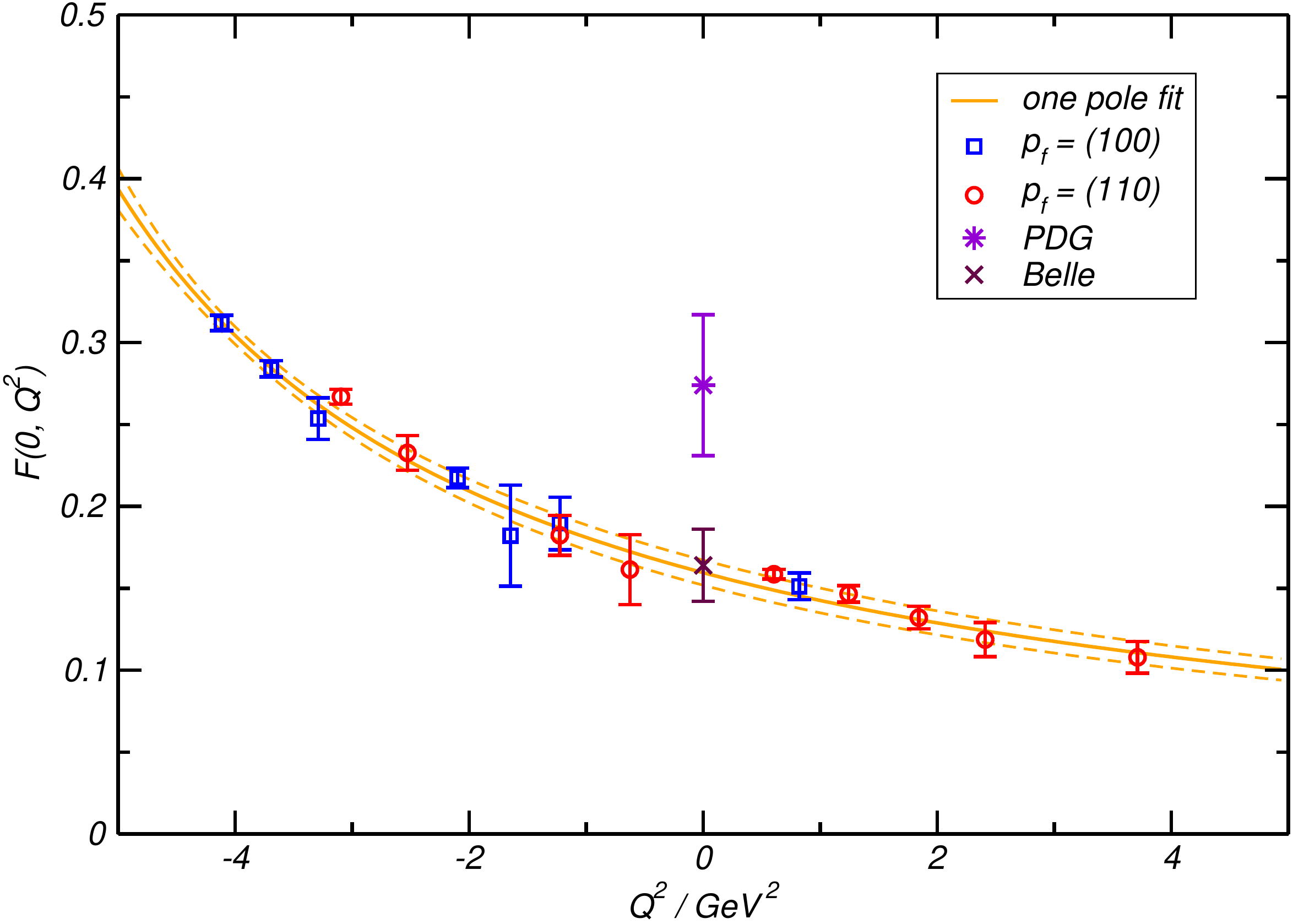} 
\includegraphics[width=75mm]{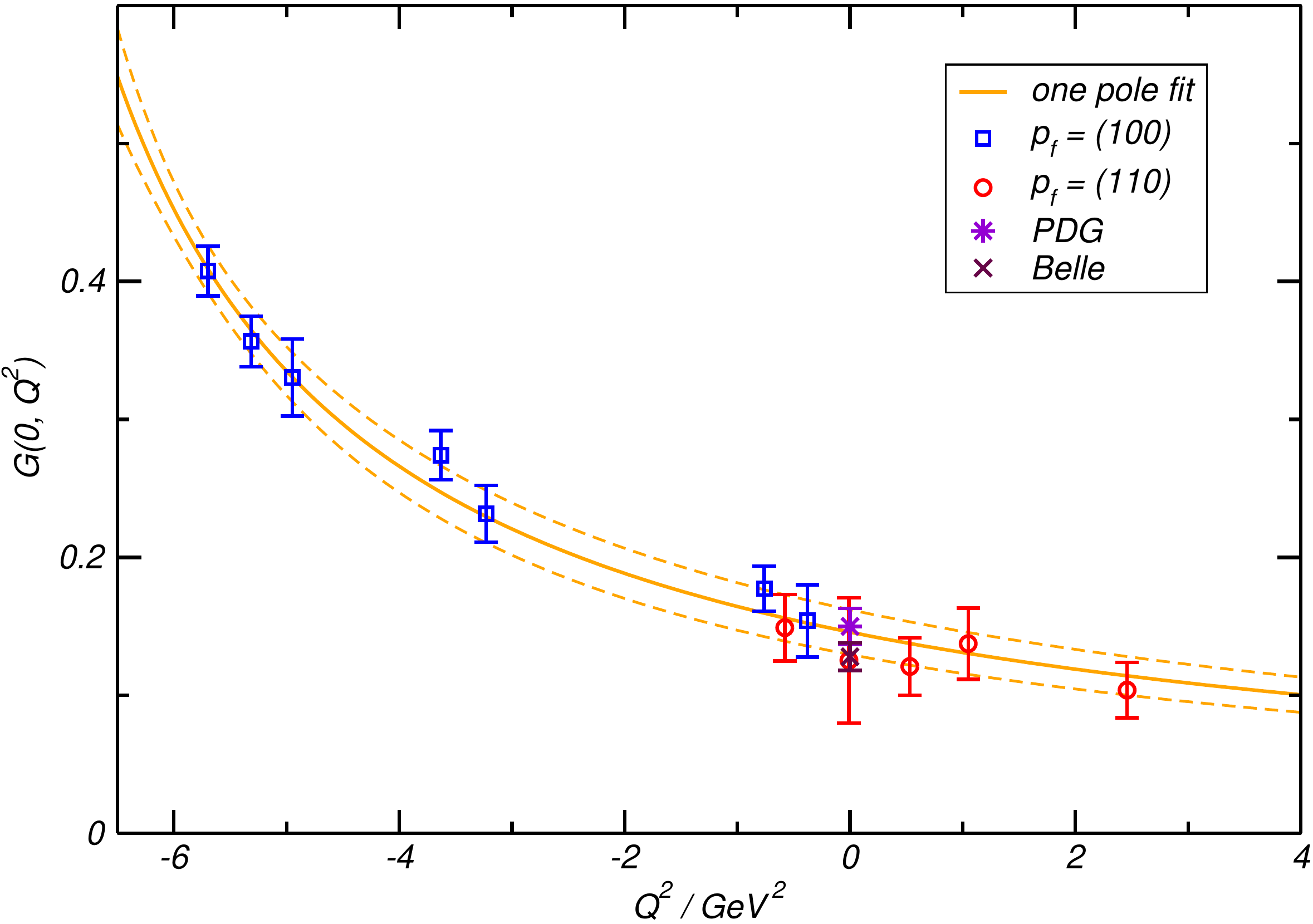} 
\end{center} 
\caption{\label{piona0}(a) $\eta_c \to \gamma \gamma^*$ amplitude. (b) $\chi_{c0}
   \to \gamma \gamma^*$ amplitude. Fits are one-pole forms as
   described in \cite{Dudek:2006ut}.} 
\end{figure}

Given the confidence that the integral can be captured on a lattice of
this temporal length, one can use a much faster method to compute the
transition form-factor that places the sum over timeslices into a
``sequential source'', reducing the computation time by a factor of
${\cal O}(L_t)$. Results using this method are shown in figure \ref{piona0}
along with PDG values and results inferred from \cite{Uehara:2007vb}.

Of course here the errors displayed on the lattice data are statistical only and must be
augmented by an uncertainty due to scaling from our fixed lattice
spacing to the continuum and one related to the lack of light-quark
loops within the quenched approximation. This is the first demonstration
of this method, such controlled studies will doubtless follow now that
efficacy has been demonstrated.

\section{Summary}

Several new techniques have emerged that much expand the range of
charmonium quantities that can be considered in lattice QCD. Initial
studies with quenched lattices are clearly systematics dominated, but
this can be expected to be improved in the near future by use of
dynamical lattices, in particular the anisotropic dynamical lattices
being generated under USQCD at Jefferson Lab. These same methods
applied to the light quark sector will provide invaluable information
for future meson spectroscopy projects like GlueX.

\begin{acknowledgments}
The research reported on in this proceedings was performed in collaboration with
Robert Edwards, David Richards and Nilmani Mathur. 

Authored by Jefferson Science Associates, LLC under U.S. DOE Contract No. DE-AC05-06OR23177. The U.S. Government retains a non-exclusive, paid-up, irrevocable, world-wide license to publish or reproduce this manuscript for U.S. Government purposes. 
\end{acknowledgments}

\bibliography{charmonium} 





\end{document}